\documentclass[aps,prd,reprint,superscriptaddress]{revtex4-2}
\usepackage[english]{babel}
\usepackage[utf8]{inputenc}
\usepackage[colorinlistoftodos, color=green!40, prependcaption]{todonotes}
\usepackage{orcidlink}
\usepackage{amsthm}
\usepackage{mathtools}
\usepackage{physics}
\usepackage{xcolor}
\usepackage{graphicx}
\usepackage{adjustbox}
\usepackage{placeins}
\usepackage[T1]{fontenc}
\usepackage{lipsum}
\usepackage{csquotes}
\usepackage{hyperref}
\usepackage{lipsum}
\usepackage[justification=raggedright, singlelinecheck=false]{caption}
\usepackage[justification=justified, singlelinecheck=false]{subcaption}

\newcommand{\n}{neutrino}

\begin{document}
\title{Gamma-Ray Echo as a Probe of Supernova Neutrino Emission Anisotropy and Long-Baseline Effects}
\author{Garv Chauhan\,\orcidlink{0000-0002-8129-8034}}
    % \email{gchauhan@vt.edu}
   \affiliation{Department of Physics, Arizona State University, 450 E. Tyler Mall, Tempe, AZ 85287-1504 USA}
\author{Cecilia Lunardini\,\orcidlink{0000-0002-9253-1663}}
    \affiliation{Department of Physics, Arizona State University, 450 E. Tyler Mall, Tempe, AZ 85287-1504 USA}
\author{Yago Porto\,\orcidlink{0000-0003-3278-0948}}
    \affiliation{Physik-Department, Technische Universit{\"a}t München, James-Franck-Stra{\ss}e, 85748 Garching, Germany}    

\begin{abstract}
During a core-collapse supernova (CCSN), the emitted electron antineutrino ($\bar{\nu}_e$) burst initiates inverse beta decay (IBD) in the outer hydrogen layer of the stellar envelope. This leads to the production of a characteristic 511 keV photon signal, termed the \emph{Gamma-Ray Echo}. The same neutrino burst, when detected on Earth in large neutrino observatories such as Hyper-Kamiokande, will offer information about the $\bar{\nu}_e$ content of the neutrino flux reaching Earth. In this work, we show that comparing the $\bar{\nu}_e$ content inferred from the gamma-ray echo with that inferred from terrestrial neutrino observations during a nearby CCSN event can offer a promising probe of large neutrino emission anisotropies and long-baseline propagation effects. In the latter case, the combined SN–Earth observations effectively act as a near–far detector configuration for neutrino propagation over an astrophysical baseline. We find that, for a near-Earth supernova, anisotropies or non-standard propagation effects at the level of tens of percent can be tested using gamma-ray telescopes with effective areas of $\mathcal{O}(10^4)~\mathrm{ cm^2}$. Therefore, our results motivate the development of the next generation of large effective area gamma-ray telescopes operating in the MeV gap. 
\end{abstract}

\maketitle

\section{Introduction}
A core-collapse supernova (CCSN) is a rich laboratory for multimessenger astrophysics, whose importance in the current era cannot be overstated. Neutrinos  carry the most direct information on the processes occurring deep inside the star, near the newly formed proto-neutron star, from where they escape immediately after core collapse, propagating directly to Earth without absorption. Likewise, gravitational waves (GWs) offer complementary information on the structure and evolution of proto-neutron star and surrounding regions~\cite{LIGOScientific:2019ryq}.
The main electromagnetic (EM) signal, instead, originates near the surface of the star at the time of shock break-out, which takes place typically hours (depending on the progenitor radius and the shockwave velocity) after the neutrino burst and GW emission.
Future observations of a galactic supernova are therefore expected to constitute a synergistic multimessenger program, where the early messengers (neutrino and gravitational waves) and the late ones (photons) test different physics of a supernova, and their connection is only indirect. 
Recent proposals, however, suggest that, for a near-Earth supernova (distance $D\lesssim 1$ kpc) a characteristic electromagnetic signal may also be observable early, starting in time coincidence with the neutrino burst and developing over tens of minutes, therefore bridging the gap between the early and late regimes~\cite{Lunardini:2023ilg}. Such a signal arises when the primary neutrino flux traverses the outer layers of the progenitor and initiates Inverse Beta Decay (IBD); the resulting IBD daughter positrons subsequently annihilate and produce a characteristic 511 keV  photon signal~\cite{BisnovatyiKogan:1975,Ryazhskaya:1999,Lu:2007wp}, which constitutes a gamma-ray echo of the neutrino burst. In the literature so far, the echo has been modeled under the simplest assumptions of spherical symmetry, and its phenomenology has not been fully explored. 

\begin{figure}[t]
    \centering
    \includegraphics[width=\linewidth]{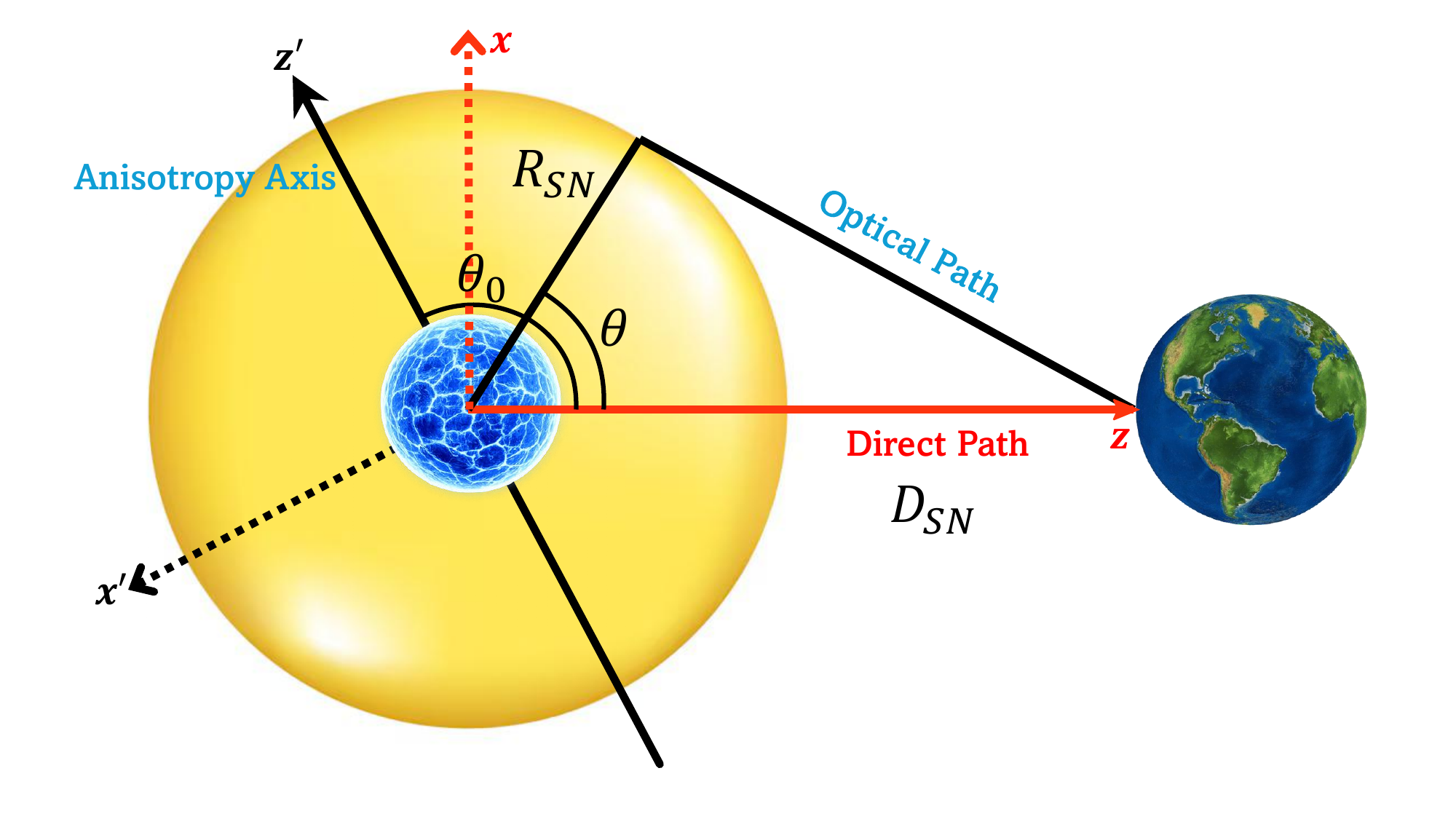}
    \caption{Schematic illustration of the gamma-ray echo geometry for anisotropic CCSN neutrino emission. The $z$-axis points along the direct SN--Earth line of sight, while the $z'$-axis denotes the neutrino-emission anisotropy axis. The polar angle $\theta$ is measured relative to the $z$-axis, and the two frames are separated by $\theta_0$. Neutrinos follow the direct path to Earth, whereas GRE photons produced in the envelope propagate along optical paths corresponding to different $\theta$.}
    \label{fig:mainfigure}
\end{figure}

Realistically, the deviation from spherical symmetry in the \n\ emission can be significant~\cite{Choi:2025igp}. It may arise from several mechanisms, including the Lepton-Number Emission Self-sustained Asymmetry (LESA)~\cite{Tamborra:2014aua,Tamborra:2014hga,Glas:2018vcs}, the Standing Accretion Shock Instability (SASI)~\cite{Tamborra:2013laa,Lin:2019wwm}, extremely strong magnetic fields~\cite{Duan:2004nc}, rapid rotation and the associated non-axisymmetric or quadrupolar emission patterns~\cite{Takiwaki:2017tpe,Walk:2019ier,Pajkos:2025zob}, magnetorotational dynamics with polar or jet-like explosion geometries~\cite{Kuroda:2020bdq} as well as 
progenitor asphericities~\cite{Muller:2017hht} and the asymmetric accretion flows 
they seed~\cite{Janka:2024xbp}. Anisotropies may also be amplified by angle-dependent flavor conversion from collective neutrino oscillations~\cite{Esteban-Pretel:2007all,Duan:2010bg}.The anisotropy in \n\ emission caused by these effects can influence the time evolution of the gamma-ray echo \cite{Lunardini:2023ilg}. 

In addition to probing stellar physics, the echo can be affected by physics beyond the Standard Model.
Indeed, the stellar envelope and neutrino detectors on Earth are analogous to a near-far detector configuration that can test long baseline neutrino physics \cite{Lunardini:2023ilg}, like oscillation into sterile species,  neutrino decay, and exotic absorption effects.
%and non-standard production channels.
These new physics phenomena acting over the very long propagation baseline can generate a mismatch between the $\bar{\nu}_e$ flux inferred from the gamma-ray echo and that inferred at terrestrial neutrino detectors, in a manner that can mimic the effect of intrinsic emission anisotropy.

In this work, we develop the phenomenology of the gamma-ray echo. We extend the formalism to the case of anisotropic neutrino emission, and develop the idea of the echo as a test of extremely long baseline neutrino physics. Specifically, we propose 
that combining observations of the echo with the neutrino signal detected on Earth allows us to extract information on the anisotropy of neutrino emission. 
We focus primarily on the anisotropy aspect, while the near-far detector concept is outlined briefly and will be developed in detail in a companion follow-up work.

This paper is organized as follows. In Section~\ref{sec:GRE}, we introduce the gamma-ray echo (GRE)
%concept, its calculation,
and generalize its formalism to anisotropic neutrino emission.
%scenarios, along with the resulting 
GRE flux and event rates are computed for these different cases. In Section~\ref{sec:HKIceCube}, we discuss the neutrino signal observed at Earth in Hyper-Kamiokande (Hyper-K)~\cite{Hyper-Kamiokande:2021frf}. In Section~\ref{sec:IBDratio}, we define a ratio between the GRE event rate observed in gamma-ray telescope to the IBD event rate measured in neutrino observatories, and discuss its dependence on the anisotropy parameter and implications for the required effective area for the 511 keV gamma-ray telescope, as well as the resulting constraints on neutrino emission anisotropies arising from either dipole or quadrupole emission. In Section~\ref{sec:BSM}, we briefly outline the long-baseline idea by considering the effect of invisible neutrino decay on the observations and assess the sensitivity of our method. Finally, we conclude in Section~\ref{sec:conclusion} with an outlook for future work.

\section{Gamma-Ray Echo and Anisotropy}
\label{sec:GRE}

\subsection{Formalism}
\label{subsec:setup} 
The concept of gamma-ray echo was first proposed by Bisnovatyi-
Kogan et al. (1975)~\cite{BisnovatyiKogan:1975} and further studied by Ryazhskaya (1999)~\cite{Ryazhskaya:1999}, when it was proposed that outgoing SN $\bar{\nu}_e$ flux passing through the outer stellar envelope dominantly composed of hydrogen will initiate inverse-beta decay (IBD) events. The final state positrons from IBD consequently undergo rapid thermalization and eventually annihilation at rest, leading to a characteristic gamma-ray signal at 511 keV :
\begin{equation}
\quad\bar{\nu}_e+p\xrightarrow{\text{IBD}}n+e^+; \quad e^+ + e^- \xrightarrow{\text{ann. at rest}} 2\gamma\,(511 \text{keV})
\end{equation}
The modern detailed treatment for the echo signal generation, luminosity curve, and detection prospects have been covered in Refs.~\cite{Lu:2007wp,Lunardini:2023ilg}. 
Here we follow the formalism of Ref.~\cite{Lunardini:2023ilg} and generalize it to the case of anisotropic \n\ emission. 

The general expression for the 511 keV signal at Earth is given by 
\begin{widetext}
\begin{equation}
    \Phi_{511}(t)=\frac{\eta_\gamma}{8 \pi D_\text{SN}^2}\frac{Y_p}{Y_e\,\sigma_c}\int_0^1\,d(\cos\theta)\int^{2\pi}_0
\,\frac{d\phi}{2\pi}\int^{\infty}_{E^\text{th}_\text{IBD}}\,{dE_\nu}\,\sigma_\text{IBD}(E_\nu)\frac{dL_\nu}{dE_\nu}\left(\theta,\,\phi,\,t_R=t-\frac{R_\text{SN}}{c}(1-\cos\theta)\right)\Theta(t_R)\,,
\label{eq:MainGREFlux}
\end{equation}
\end{widetext}
where $D_\text{SN}$ is the distance to the SN progenitor, $dL_\nu/dE_\nu$ is the differential number luminosity, $\sigma_{c}$ is the Compton cross-section at 511 keV, $\sigma_\text{IBD}$ is the IBD cross-section dependent on the incoming neutrino energy $(E_\nu)$ and IBD threshold energy $E^\text{th}_\text{IBD}$~\cite{Vogel:1999zy,Strumia:2003zx}. $Y_p$ is the free-proton (hydrogen) fraction available as IBD target and $Y_e$ is the total number of atomic electrons per baryon at the stellar envelope edge. We use $Y_p=0.7$ and $Y_e=0.85$ as fiducial values in this work~\cite{Lu:2007wp,Lunardini:2023ilg}. $\eta_\gamma\approx 1.74$ is the effective number of 511 keV photons produced per positron~\cite{Lu:2007wp,Lunardini:2023ilg}. The variables $(\phi,\theta)$ denote the azimuthal angle and polar angle respectively; see fig. \ref{fig:mainfigure} for an illustration of the geometry of the system. We define the $z$-axis of our coordinate system as the line-of-sight (LOS) from SN progenitor to the observer at Earth. The $x-y$ plane lies in the plane perpendicular to the LOS (projection of the stellar surface facing the observer). $t$ denotes the time elapsed on Earth since the detection of first neutrino event and the factor ${R_\text{SN}}(1-\cos\theta)/{c}$ accounts for the delay of the arrival at Earth of the echo photons produced at a generic point on the stellar surface, with respect to photons that are produced and propagating along the LOS (fig. \ref{fig:mainfigure}).
Our expression reproduces the result derived in Ref.~\cite{Lunardini:2023ilg}, in the limit of spectrum-averaged $\sigma_{\text{IBD}}$ and isotropic \n\ emission. 

Since the initial outgoing neutrino flux is produced near the PNS at radial distance far smaller than $ R_{\text{SN}}$ and the luminosities used in this work are being used for calculating effects at larger distances, we can assume them to be viewing angle-averaged~\cite{Tamborra:2014hga}. 
For studying the anisotropic scenario, we consider the following functional form for the neutrino luminosity $L_\nu(\theta',r',t)$ in polar coordinates with a residual dipole (see fig. \ref{fig:mainfigure})
\begin{equation}
    L_\nu(\theta',\phi',t) = L^\text{iso}_\nu(t)\,(1+\epsilon_D\cos{\theta'})~.
    \label{eq:LnuExp1}
\end{equation}
where $L^\text{iso}_\nu(t)$ is the isotropic (or equivalently the monopole component)  luminosity independent of anisotropy, $\epsilon_D \in[-1,1]$ is the dipole anisotropy parameter, $\theta'$ is defined as the polar angle measured from the dipole axis (chosen to be the $z'$-axis, see Fig. \ref{fig:mainfigure}). 
For simplicity, here we assume $\epsilon_D$ to be time- and energy-independent. Therefore, the energy dependence of the luminosity in Eq.~\eqref{eq:MainGREFlux} arises from $L^\text{iso}_\nu(t)$. Note that $L^\text{iso}_\nu(t)$ also retains the temporal evolution of the neutrino luminosity as observed in the SN simulations. In addition, we consider only the case with no precession of the anisotropy axis about the rotation axis. Our results are also applicable to cases with no stellar rotation or with a small precession angle. We focus on these cases because a precessing anisotropy axis can produce a neutrino lighthouse effect, allowing the anisotropy to be resolved directly through modulation of the IBD event rate in neutrino observatories~\cite{Takiwaki:2017tpe}. By contrast, in the absence of stellar rotation or appreciable precession, the neutrino signal at Earth arrives along only a single viewing direction and is therefore degenerate with the unknown monopole neutrino luminosity. In these cases, the GRE provides an excellent way to break this degeneracy by providing a complementary average over the observer-facing hemisphere.

For reference values of $\epsilon_D$, a relevant observable from time-dependent simulations is provided by the time-integrated momentum-asymmetry parameter. The values reported in Ref.~\cite{Janka:2024xbp} correspond to effective dipole anisotropy $|\epsilon_D|$ up to $0.11$ over the simulated $0.25$--$4.1\ {\rm s}$ interval, with the largest values occurring in low-mass, LESA-dominated models. However, this comparison is approximate because their reported observable is energy-momentum-weighted rather than IBD-weighted and the simulations do not cover the complete cooling signal.

Generally, the dipole $z'$-axis and the $z$-axis do not coincide, therefore we need to rewrite the anisotropic luminosity in the gamma-ray echo (un-primed) frame. To simplify our analysis, we can set the $y$-axis of our gamma-ray echo coordinate frame to be aligned with the $y'$-axis of the dipole. Therefore, the two coordinate systems differ by a rotation in the $x$-$z$ plane (about the $y$-axis). Let us denote this angle between the $z'$-axis and the $z$-axis to be $\theta_0 \in[0,\pi]$ (see Fig. \ref{fig:mainfigure}), 
and use the angle conversion: 
\begin{equation}
\cos{\theta'}=\sin{\theta_0}\,\sin{\theta}\cos{\phi}+\cos{\theta_0}\cos{\theta}~.
    \label{eq:lum}
\end{equation}
We will use the above expressions in Eqs.~\eqref{eq:LnuExp1} and ~\eqref{eq:lum} for the differential luminosity with a net dipole in the study of anisotropic neutrino emission. Note that the angular average over the azimuthal angle does not wash away the dependence on the dipole anisotropy parameter in Eq.~\eqref{eq:MainGREFlux}
\begin{align}
\int^{2\pi}_0
\,\frac{d\phi}{2\pi}\,\left[1+\epsilon_D\,(\sin{\theta_0}\,\sin{\theta}\cos{\phi}+\cos{\theta_0}\cos{\theta})\right] \nonumber \\= 1 +\epsilon_D\,\cos{\theta_0}\cos{\theta}\,.
\end{align}
Upon further integration of the azimuthally averaged rate over the polar angle, the total GRE event count for the dipole case ${N}_{\gamma}^{D}$ reduces to the following form,
\begin{align}
{N}_{\gamma}^{D}&=\int_0^{\infty} dt\,\,A_{\text{eff}}\,\Phi_{511}(t)\\\,&={N}_{\gamma}^{0}\,\left(\int_0^1\,d(\cos\theta)
\,(1+\epsilon_D\cos{\theta_0}\cos{\theta})\right) \nonumber\\ &= {N}_{\gamma}^{0}\,\left(1+\frac{\epsilon_D\cos{\theta_0}}{2}\right)\, .
\label{eq:NestimateDipole}
\end{align}
where $A_{\text{eff}}$ is the effective area of the telescope at 511 keV and ${N}_{\gamma}^{0}$ is the total GRE event count in the isotropic case. In an analogous fashion, we extend our framework to account for quadrupolar emission asymmetry. The neutrino luminosity $L_{\bar{\nu}_e}(\theta', r', t)$ with a residual quadrupole component is parameterized via the second Legendre polynomial,
\begin{align}
    L_\nu(\theta',\phi',t) &= L^\text{iso}_\nu(t)\,[1+\epsilon_Q\, P_2(\cos{\theta'})]\,, \nonumber \\
    P_2(\cos{\theta'})&=\left(\frac{3\cos^2{\theta'}-1}{2}\right)\, . 
    \label{eq:Lnu2}
\end{align}
where $\epsilon_Q \in[-1,2]$ is the quadrupole anisotropy parameter and $\theta'$ is defined as the polar angle measured from the quadrupole axis (chosen to be the rotation-axis). 

We note that, as in the dipole case, the angular average over the azimuthal angle does not wash away the dependence on the parameter $\epsilon_Q$. However, after polar integration, the time-integrated event rate for the quadrupole case ${N}_{\gamma}^{Q}$ is independent of $\epsilon_Q$.
\begin{widetext}
\begin{align}
{N}_{\gamma}^{Q}=\int_0^{\infty} dt\,\,A_{\text{eff}}\,\Phi_{511}(t)&={N}_{\gamma}^{0}\,\left( \int_0^1\,d(\cos\theta)\int^{2\pi}_0 
\,\frac{d\phi}{2\pi}\,\left[1+\frac{3\,\epsilon_Q}{2}\,(\sin{\theta_0}\,\sin{\theta}\cos{\phi}+\cos{\theta_0}\cos{\theta})^2-\frac{\,\epsilon_Q}{2}\right]\right)  \nonumber \\ &= {N}_{\gamma}^{0}\,\left(\int_0^1\,d(\cos\theta)
\,\left[1+\epsilon_Q\,P_2(\cos{\theta_0})\,P_2(\cos{\theta})\right]\right) = {N}_{\gamma}^{0}\, .
\label{eq:NestimateQuadrupole}
\end{align}
\end{widetext}
The above results can also be understood in a geometric manner. For dipolar emission, depending on the orientation of the dipole axis, the observer-facing hemisphere contains a larger contribution from either the hotspot or the coldspot. Averaging the dipolar pattern over this hemisphere reduces the projected anisotropy $\epsilon_D \cos\theta_0$ by a factor of two, leading to $N_{\gamma}^{D} \propto 1+\epsilon_D\cos\theta_0/2$. For quadrupolar emission, however, the enhancement and suppression regions balance each other within the observer-facing hemisphere. The quadrupolar contribution therefore cancels upon angular integration, and the total integrated event count $N_{\gamma}^{Q}$ is independent of $\epsilon_Q$.

Recent axisymmetric calculations of extremely rapidly rotating progenitors find IBD-dominated pole-to-equator detector-count ratios of approximately 1.5 during the bounce phase and up to 2.5 during the accretion phase~\cite{Pajkos:2025zob}; if represented by a pure $P_2\,(\cos{\theta'})$ pattern, these ratios would correspond to $\epsilon_Q\simeq0.3$ and $0.7$, respectively. Although these results are reported for only the first $300\ {\rm ms}$, they provide phenomenological motivation for the values of $\epsilon_Q$ considered in our work.

\begin{table*}
    \centering
    \begin{tabular}{|c|c|c|c|c|c|c|c|c|}
     \hline
      Candidate  & Distance (kpc) & Radius (cm)  & Initial/ZAMS Mass  & Galactic & Galactic & $\Phi_{gal}$ & ${N}_{\gamma}$ \\ 
      & & & ($M_\odot$) & Longitude ($^\circ$) &  Latitude ($^\circ$) & ($\text{ cm}^{-2} \text{ s}^{-1}$) &  \\ \hline
      Rigel   &  $0.264\pm0.024$ & $5.5\times10^{12}$ & $24.0\pm{3.0}$ & 209.24 & -25.24& $2.9\times10^{-8}$ & 158\\
      $\epsilon$ Pegasi  & $0.211\pm 0.006$ &$1.3\times10^{13}$ & $11$-$12$ &65.57 & -31.45& $7.5\times10^{-8}$ & 104 \\
    Betelgeuse  & $0.222\pm0.040$ & $5.3\times10^{13}$ & $18$-$21$ & 199.78 & -8.95 & $2.4\times10^{-7}$ & 224\\
    \hline
    \end{tabular}
    \caption{The estimated distance, radius, initial mass, galactic coordinates~\cite{Przybilla:2005na,Moravveji:2012,vizier:2018,Mukhopadhyay:2020ubs,Joyce:2020,Neuhauser:2022}, the expected diffuse gamma-ray galactic background at 511 keV~\cite{Skinner:2014,Siegert:2015knp} and the integrated 511 keV GRE photon count for the nearby SNe progenitor candidates discussed in this work, assuming isotropic neutrino emission. The total photon counts and  background rate are quoted for a gamma-ray telescope with $A_\text{eff} = 10^4~\text{cm}^2$   and angular resolution of $3^\circ$. }
    \label{tab:SNcandidates}
\end{table*}

\subsection{Gamma-ray detection}
\label{subsec:gammadetect} 
Since the predicted event count depends sensitively on the distance to the SN progenitor, we focus on the nearest candidate progenitors~\cite{Lunardini:2023ilg}. The relevant source parameters --- distance, progenitor radius, mass and galactic coordinates --- for the three candidates considered in this work, namely Rigel, Betelgeuse, and $\epsilon$~Pegasi, are summarized in Table~\ref{tab:SNcandidates}~\cite{Mukhopadhyay:2020ubs}. For the GRE signal to be observable, it should be stronger than the relevant astrophysical backgrounds. In this case, the dominant contribution arises from diffuse 511 keV gamma-rays produced by positron annihilation in the Galactic disk~\cite{Skinner:2014,Siegert:2015knp}. In Ref.~\cite{Lunardini:2023ilg}, the estimated background was calculated using observations reported in Ref.~\cite{Skinner:2014}. We calculate the expected Galactic background using the two-dimensional model for the diffuse disk emission of Ref.~\cite{Siegert:2015knp}, with a total disk line flux $F=1.66\times10^{-3}\text{ cm}^{-2}\text{ s}^{-1}$ and Gaussian disk widths of $60^\circ$ in longitude and $10.5^\circ$ in latitude. We note that the analysis of the SPI observations in Ref.~\cite{Siegert:2015knp} finds a thicker disk model to provide a better fit to the data than the thinner disk morphology reported in the earlier analysis of Ref.~\cite{Skinner:2014}. We provide the resulting refined estimates of $\Phi_{gal}$ in the second last column of Table~\ref{tab:SNcandidates}, assuming a detector angular resolution of $3^\circ$.

Currently, the next-generation gamma-ray telescope COSI~\cite{Tomsick:2021wed}, planned for launch in 2027 and designed to improve line sensitivity at 511~keV, will achieve an effective area of only $\sim 20~\text{cm}^2$ at this energy. The proposed large effective-area concepts e-ASTROGAM~\cite{e-ASTROGAM:2017pxr} and AMEGO~\cite{Kierans:2020otl} represent a significant step forward, with projected effective areas of up to $\sim 3000~\text{cm}^2$. There is also a recent proposal for  APT~\cite{Buckley2019Advanced}, which will reach impressive effective areas of $A_{\text{eff}} = 10^4~\text{cm}^2$, a threshold that we identify in this work as necessary to constrain the $\bar{\nu}_e$ fraction --- and the neutrino emission anisotropy --- to approximately $|\epsilon_D|=0.15$ and $|\epsilon_Q|=0.18$ at $68\%$~C.L. during a nearby galactic CCSN. This highlights a clear and pressing need for dedicated next-generation MeV gamma-ray instrumentation with substantially larger collection areas than those currently planned.

\begin{figure*}[ht!]
    \centering
    \hspace{0.7in}
        \begin{subfigure}[t]{0.48\linewidth}
        \includegraphics[width=\linewidth]{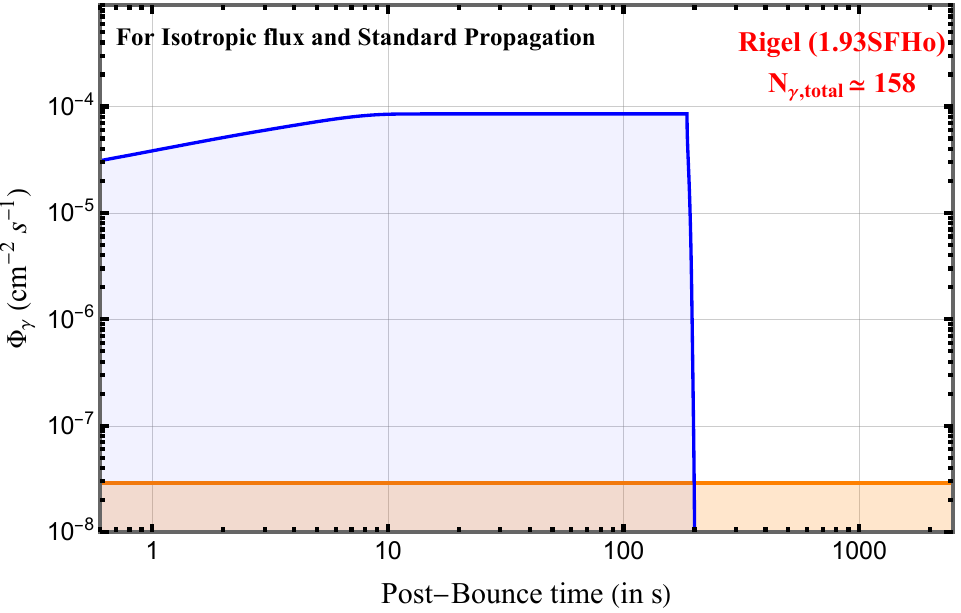}
        \caption{}
        \label{fig:GREexp_c}
    \end{subfigure}
    \newline
    \begin{subfigure}[t]{0.48\linewidth}
        \includegraphics[width=\linewidth]{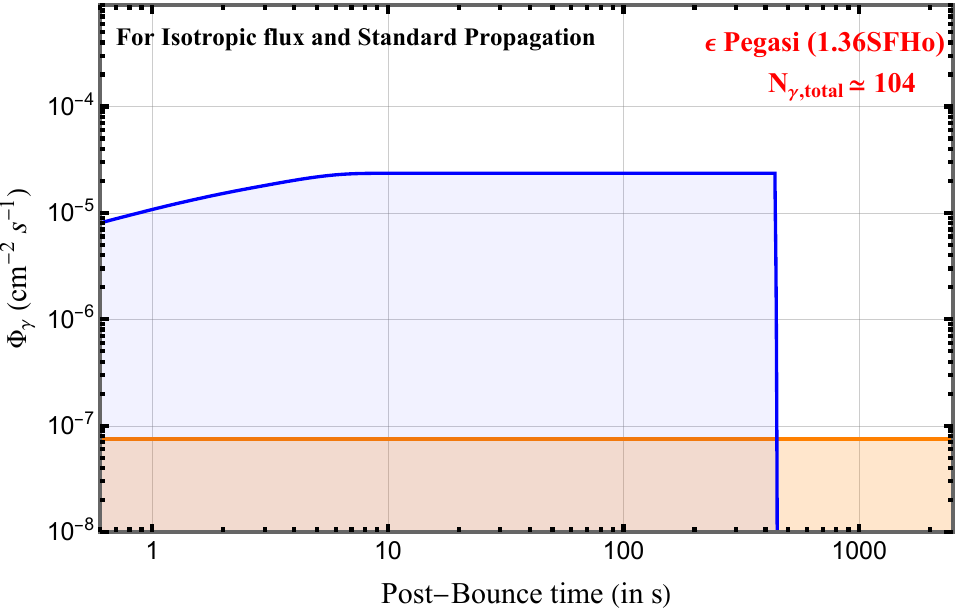}
        \caption{}
        \label{fig:GREexp_b}
    \end{subfigure}
    \hfill
    \begin{subfigure}[t]{0.48\linewidth}
        \includegraphics[width=\linewidth]{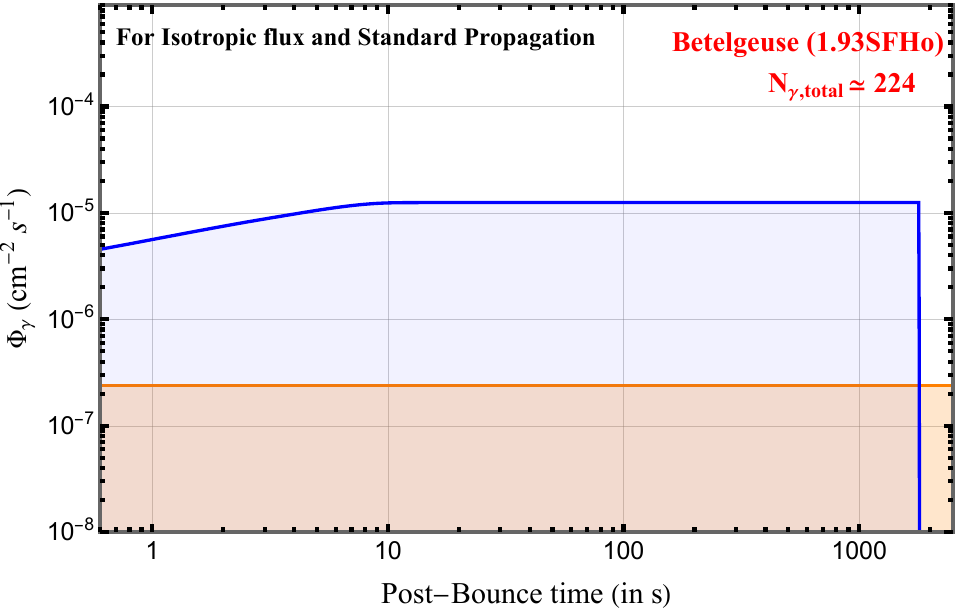}
        \caption{}
        \label{fig:GREexp_d}
    \end{subfigure}
    \caption{The Gamma-Ray Echo flux as a function of post-bounce time for isotropic neutrino emission under standard propagation (blue lines), shown for three SN progenitors:  \textbf{(a)} Rigel, \textbf{(b)} $\epsilon$~Pegasi, and  \textbf{(c)} Betelgeuse, respectively. 
    The total photon count is quoted for $A_\text{eff} = 10^4~\text{cm}^2$. The expected diffuse gamma-ray galactic background at 511 keV is shown in orange.}
    \label{fig:GREexp}
\end{figure*}

As the reference neutrino emission profiles for our calculations, we adopt two isotropic neutrino luminosity spectra $L_{\bar{\nu}_e}(t)$ from the one-dimensional CCSN simulations performed by the Garching group using the SFHo equation of state (EoS)~\cite{Steiner:2012rk, SNprofiles:Garching}, carried out with the \texttt{PROMETHEUS-VERTEX} code~\cite{Hudepohl:2009tyy}. Specifically, for $\epsilon$~Pegasi we adopt the $9~M_\odot$ progenitor model (1.36SFHo), while for Betelgeuse and Rigel we adopt the $20~M_\odot$ progenitor model (1.93SFHo)~\cite{Fiorillo:2023frv}. We select the representative model for each candidate by comparing its progenitor ZAMS mass with the inferred initial mass of the star, rather than with its present-day mass (see Table~\ref{tab:SNcandidates}). The neutrino flux and spectra are provided for a resting observer frame at 500 km, in terms of pinched-thermal (alpha) spectral parameters  characterized by the mean energy $\langle E_{\bar{\nu}_e} \rangle$, a second energy moment $\langle E_{\bar{\nu}_e}^2 \rangle$ and total neutrino energy luminosity~\cite{Keil:2002in,Tamborra:2012ac}. 

It is worth emphasizing that the interpretation of the gamma-ray echo is largely insensitive to the details of flavor conversion, since our analysis is focused on characterizing the normalization of the $\bar{\nu}_e$ flux at the stellar surface itself. The relevant $\bar{\nu}_e$ flux that initiates IBD in the hydrogen envelope undergoes no further flavor evolution between the stellar surface and Earth~\cite{Dighe:1999bi}. Nevertheless, for internal consistency of our setup, we assume throughout that the simulated neutrino fluxes have undergone standard MSW conversion in the normal mass ordering (NH). Therefore, neutrino fluxes at the stellar surface are linear combinations of the initial fluxes $L_{\bar{\nu}_e}^0$ predicted by the SN simulations~\cite{Dighe:1999bi}. 
\begin{equation}
  L_{\bar{\nu}_e} = \cos^2\theta_{12}\,\cos^2\theta_{13}\,L_{\bar{\nu}_e}^0   + (1-\cos^2\theta_{12}\,\cos^2\theta_{13})\,L_{\bar{\nu}_x}^0\, ,
\end{equation}
where $x$ refers to either of the $\mu$ or $\tau$ flavor neutrino flux and, $\cos^2\theta_{12}=0.693$ and $\cos^2\theta_{13}=0.9783$ are the vacuum neutrino mixing parameters~\cite{ParticleDataGroup:2026mpi}. Adopting an alternative oscillation scenario would modify the total predicted event counts only mildly, without affecting the qualitative conclusions of our analysis.

In Figs.~(\ref{fig:GREexp_c},\ref{fig:GREexp_b}, \ref{fig:GREexp_d}), we show the GRE flux and the total expected event rate for three SN progenitors, assuming isotropic neutrino emission and standard propagation. The instantaneous GRE flux can reach up to $\mathcal{O}(10^{-4})\text{ cm}^{-2}\text{ s}^{-1}$ and the diffuse-background contribution remains subdominant in all cases. It can also be seen that the characteristic timescale of the GRE emission is strongly controlled by the stellar radius $R_{ SN}$. Therefore, for a given neutrino-emission model and the same distance to the progenitor, a candidate with a smaller radius will produce a higher instantaneous GRE flux. Since the remnant neutron star (NS) mass correlates with the overall strength of the neutrino emission, the expected event rate for $\epsilon$-Pegasi is the lowest among the three, despite it being the nearest progenitor. For Rigel and Betelgeuse, the underlying SN neutrino flux is identical given that both are modeled with the same progenitor mass; however, since Rigel is the more distant of the two, its total predicted photon count is correspondingly lower. 

\begin{figure*}[htbp!]
    \centering 
        \includegraphics[width=0.49\linewidth]{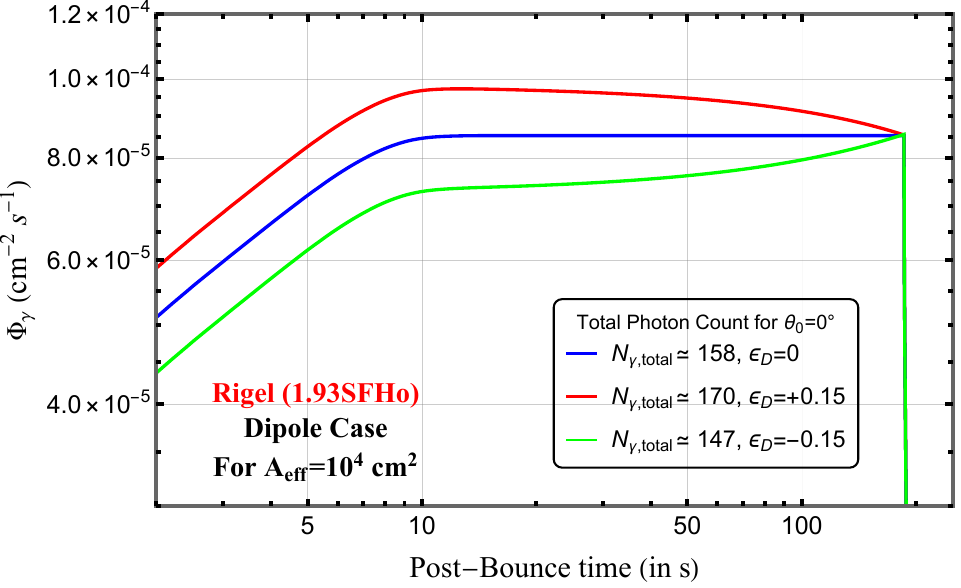}
        \includegraphics[width=0.49\linewidth]{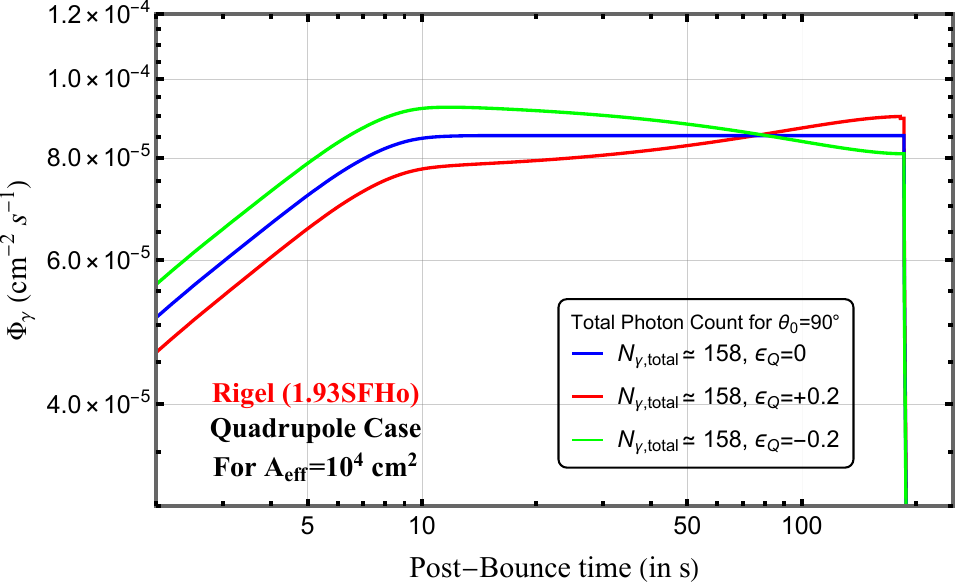}
    \caption{The GRE lightcurves for Rigel for different values of the anisotropy parameters $\epsilon_D$ (left) and $\epsilon_Q$ (right). The case for isotropic emission corresponds to $\epsilon_{D}=0$ and $\epsilon_{Q}=0$, for neutrino luminosity profile taken from 1.93SFHo simulation for fixed $\theta_0=0^\circ\,(90^\circ)$. The total photon counts are shown assuming $A_\text{eff}=10^4\text{ cm}^2$.}
    \label{fig:GREAnisoRigel}
\end{figure*}

In Fig.~\ref{fig:GREAnisoRigel}, we present the GRE lightcurves for the case of anisotropy arising from dipolar (left panel) or quadrupolar (right panel) neutrino emission, computed using Eq.~\eqref{eq:MainGREFlux} for the case of Rigel, assuming an effective detector area of $10^4~\text{cm}^2$.  We now turn to the most distinctive features of the anisotropic emission, displayed for several values of the anisotropy parameter $\epsilon_D$ and $\epsilon_Q$. The case $\epsilon_D=0\,(\epsilon_Q=0)$ corresponds to isotropic emission. We study the light curves for fixed value of $\theta_0$. 

For dipole axis oriented  toward the terrestrial observer $\theta_0=0^\circ$, a positive value of $\epsilon_D$ (e.g., $\epsilon_D = +0.15$, red curve) signifies that the neutrino emission is enhanced at smaller polar angles. Since smaller polar angles correspond to shorter photon travel times to Earth, the GRE flux rises more steeply at early times. As the signal accumulates contributions from progressively larger polar angles --- where the emission is weaker --- the flux gradually decreases and asymptotes toward the isotropic average. This behavior is clearly seen in the red curve. By the same reasoning, a negative value of $\epsilon_D$ implies that the emission is reduced at smaller polar angles. Consequently, the GRE flux peaks at later times, as is evident from the green curve ($\epsilon_D = -0.15$). In addition, the total photon count is in excellent agreement with our estimate in Eq.~\eqref{eq:NestimateDipole}.

For quadrupole axis oriented  away from the observer $\theta_0=90^\circ$, a positive value of $\epsilon_Q$ (e.g., $\epsilon_Q = +0.2$ in the right panel) signifies that the neutrino emission is enhanced at both poles. Since for quadrupole emission enhancement and suppression regions balance each other within a hemisphere, the enhancement at the poles implies a suppression in the equatorial region. Therefore, GRE flux rises slower than the isotropic case at early times for $\epsilon_Q = +0.2$.  However,  as the signal accumulates contributions from progressively larger polar angles --- where the emission is stronger from both the polar regions --- the flux gradually increases and asymptotes towards a value larger than the isotropic average. This behavior is clearly seen in the red curve. By the same reasoning,  opposite effect occurs for a negative value of $\epsilon_Q$ that implies that the GRE flux peaks at earlier times, as is evident from the green curve ($\epsilon_Q = -0.2$) in Fig.~\ref{fig:GREAnisoRigel}. The total photon count for the quadrupole case is independent of $\epsilon_Q$ as estimated in Eq.~\eqref{eq:NestimateQuadrupole}.

For new physics scenarios in which the neutrino emission from the SN itself is isotropic, i.e.\ $L_{\bar{\nu}_e}(\theta, \phi, t) = L_{\bar{\nu}_e}(t)$, but where propagation effects along the SN-to-Earth baseline introduce non-standard attenuation or oscillation effects, we parametrize the flux arriving at Earth relative to that at the SN source via a simple rescaling parameter,
\begin{equation}
    L_{\bar{\nu}_e}^{\text{Earth}}(t) = (1 - \epsilon_{\text{BSM}})\, L_{\bar{\nu}_e}^{\text{SN}}(t),
    \label{eq:LnuBSM}
\end{equation}
where $\epsilon_{\text{BSM}}$ encapsulates the net effect of any beyond the Standard Model (BSM) physics on the propagation. We note that such en-route effects may in general be energy dependent; however, since the gamma-ray echo signal depends on the convolution of both the neutrino flux normalization and the energy spectrum, disentangling these two effects is non-trivial in practice. The simple rescaling in Eq.~\eqref{eq:LnuBSM} therefore provides a useful and model-agnostic parameter for comparing the predictions of different BSM scenarios against one another.

\section{Neutrino Signal at Earth}
\label{sec:HKIceCube}
While large effective-area $\gamma$-ray telescopes are required to observe the GRE signal, existing large-volume neutrino observatories will register thousands to millions of neutrino events during a nearby galactic CCSN, depending on the distance to the progenitor~\cite{JUNO:2015zny,DUNE:2020zfm,Hyper-Kamiokande:2021frf,IceCube:2023ogt}. In this work, we consider the upcoming water-Cherenkov detector, Hyper-K, where the dominant detection channel for neutrino event detection is through IBD. Hyper-K offers superior energy resolution at a comparably large event rate, enabling a more detailed reconstruction of the neutrino energy spectrum~\cite{Hyper-Kamiokande:2021frf}.

The IBD event rate in Hyper-K initiated by the $\bar{\nu}_e$ spectrum arriving at Earth, is given by
\begin{widetext}
    \begin{equation}
    \mathcal{R}_\text{HK}(t)=
    \frac{N_p}{4\pi D_\text{SN}^2}\int^{\infty}_{E^\text{th}_\text{IBD}}\,{dE_\nu}\,\sigma_\text{IBD}(E_\nu)\frac{dL_\nu}{dE_\nu}
     \left(\theta=0,\phi,\theta_0,t\right)\,\varepsilon_{\rm eff}(E_{\nu})\, ,
\end{equation}
\end{widetext}
where $N_p$ is the total number of target protons in the 187 kton fiducial mass and $\varepsilon_{\rm eff}(E_\nu)$ is the detection efficiency~\cite{Hyper-Kamiokande:2021frf}.

\begin{figure}[t!]
    \centering \includegraphics[width=\linewidth]{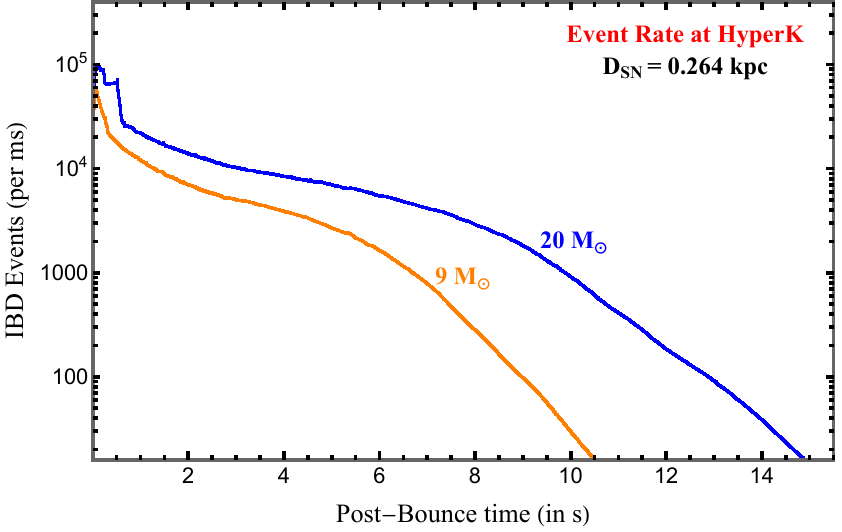}
    \caption{Event rates as a function of post-bounce time at Hyper-K for the two simulated neutrino profiles from $9$ (orange line) and $20\, M_\odot$ (blue line) SN progenitors at a distance $D_\text{SN}=0.264$ kpc.}
    \label{fig:IBDHK}
\end{figure}

The expected number of IBD events in Hyper-K for Rigel as the SN progenitor is of $\mathcal{O}(10^8)$, implying that the detected $\bar{\nu}_e$ event yield can be statistically constrained to a precision of $\mathcal{O}(10^{-4})$. In Fig.~\ref{fig:IBDHK}, we display the event rate as a function of post-bounce time at Hyper-K for the two simulated neutrino emission profiles from the $9~M_\odot$ and $20~M_\odot$ SN progenitors, assuming a Rigel-like source distance of $D_\text{SN} = 0.264$~kpc. The Hyper-K measurement is essentially background-free over the duration of the burst. The neutrino signal remains clearly detectable for such a nearby progenitor out to timescales of $10$--$15$~s post-bounce, which is sufficient to characterize the neutrino emission profile and predict the expected GRE signal for gamma-ray telescope observations. 

\section{IBD event ratio}
\label{sec:IBDratio}
\begin{figure*}[t!]
    \centering \includegraphics[width=0.49\linewidth]{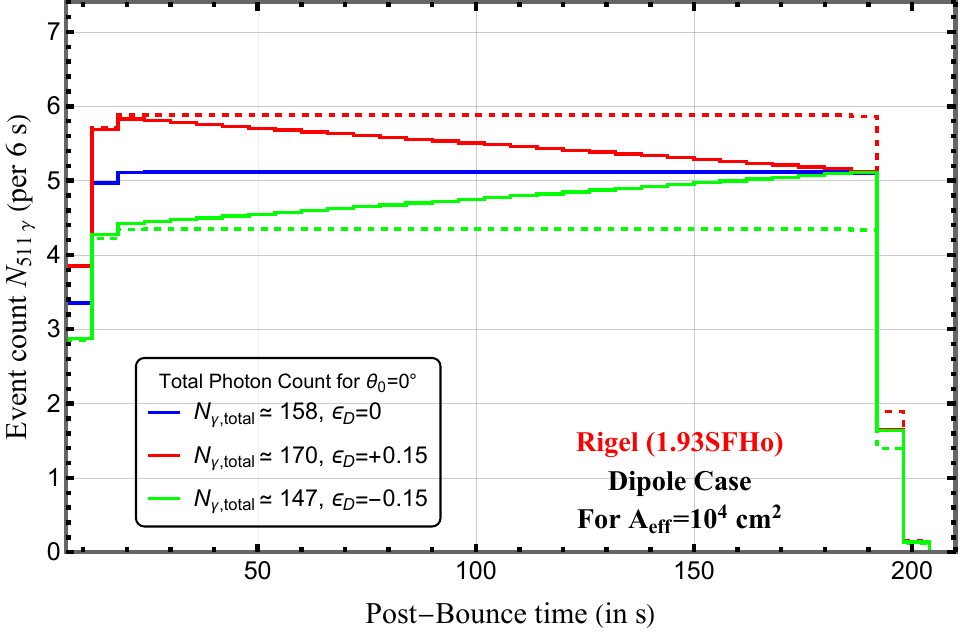}
    \includegraphics[width=0.49\linewidth]{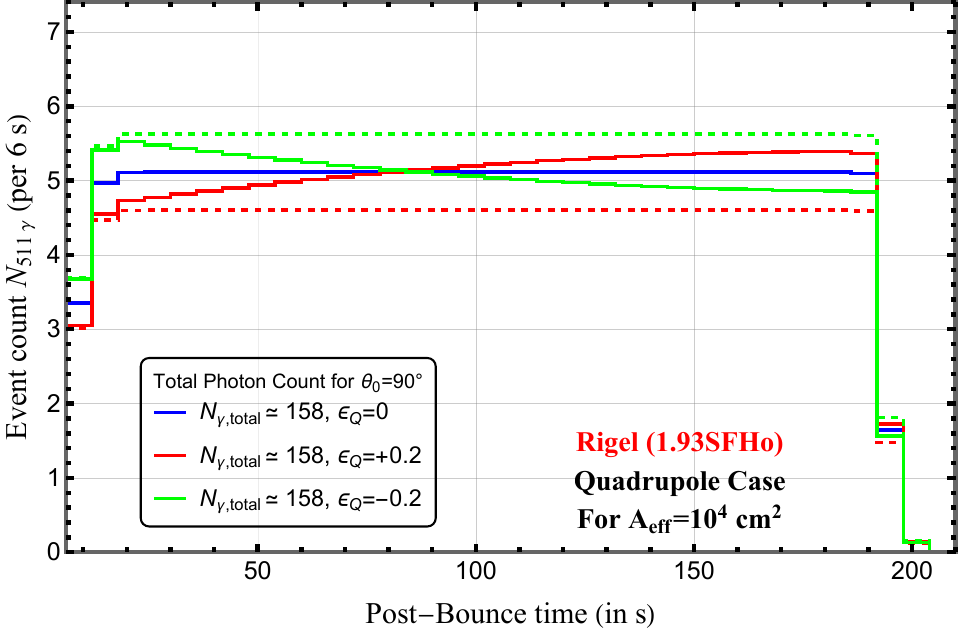}
    \caption{Expected binned photon counts in the gamma-ray detector for Rigel as a function of post-bounce time for different values of the anisotropy parameters $\epsilon_D$ (left) and $\epsilon_Q$ (right)  for $A_\text{eff}=10^4\text{ cm}^2$ and fixed $\theta_0$, assuming Rigel as the SN progenitor. Solid curves show true anisotropic GRE counts. Dotted curves show isotropic predictions inferred from the corresponding Hyper-K measurement.}
    \label{fig:EventsRigel}
\end{figure*}

\begin{figure*}[t!]
    \centering \includegraphics[width=0.49\linewidth]{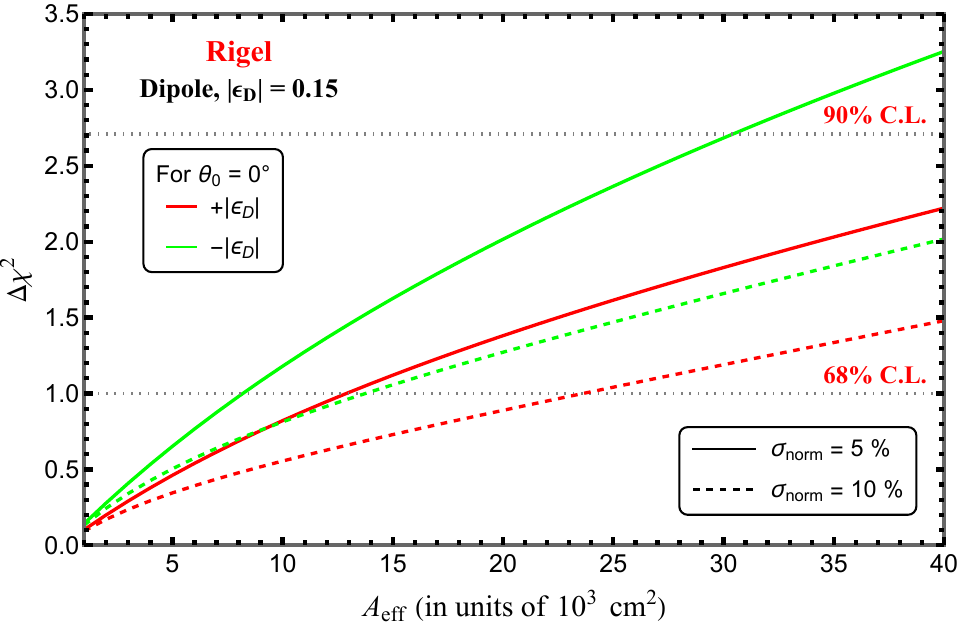}
    \includegraphics[width=0.49\linewidth]{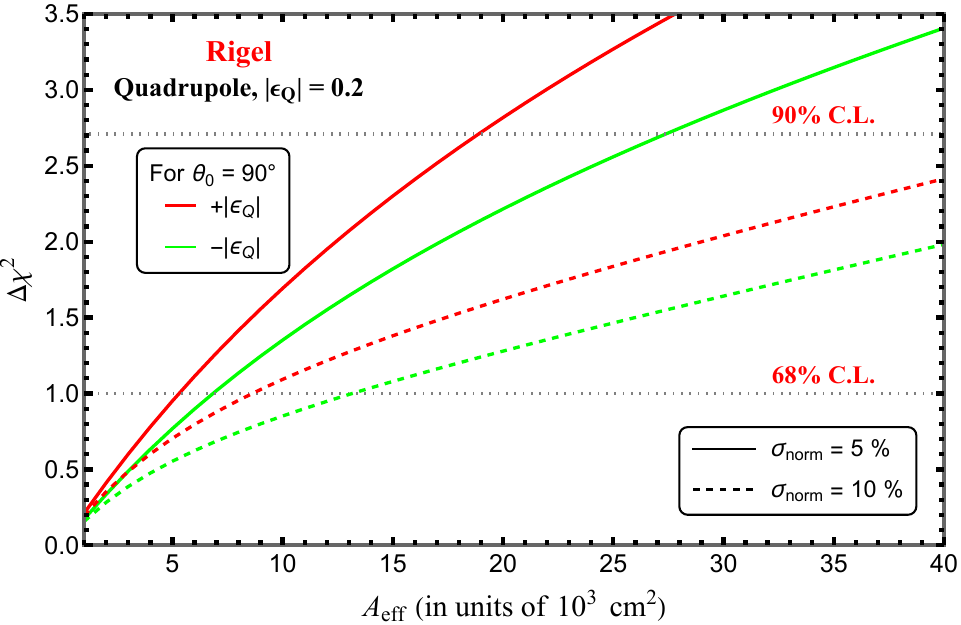}
    \caption{$\Delta\chi^2$ as function of effective area of the 511 keV gamma-ray detector for two different values of the anisotropy parameters $\epsilon_D$ (left) and $\epsilon_Q$ (right) and for two different values of $\sigma_\text{norm}=\{5\%,\,10\%\}$ uncertainty arising from the stellar systematics, assuming Rigel as the SN progenitor.}
    \label{fig:ChiSquaredRigel}
\end{figure*}
As discussed in the preceding section, the high event statistics accumulated at Hyper-K during a nearby CCSN event will allow the $\bar{\nu}_e$ flavor content of the neutrino burst arriving at Earth to be determined with great precision. In this section, we demonstrate that comparing the total IBD event count at Hyper-K against the GRE photon count provides a direct probe of the anisotropy of the SN neutrino emission. To make this precise, we define the following observable ratio of total GRE photon count to total IBD event count in Hyper-K,
\begin{align}
\text{R}^\text{observed}&=\frac{\int_0^{\infty} dt\,A_{\text{eff}}\,\Phi_{511}(t)}{\int_0^{\infty} dt\,\mathcal{R}_\text{HK}(t)}.
\end{align}
Under the null hypothesis of isotropic SN neutrino emission, this ratio can also be calculated theoretically,
\begin{widetext}
\begin{align}
\text{R}_{\textbf{iso}}^\text{theory}&
= \frac{\dfrac{\eta_\gamma\,A_{\text{eff}}}{8 \pi D^2}\dfrac{Y_p}{Y_e\,\sigma_c}\int_0^{\infty} dt\,\int_0^{1}\,d(\cos\theta){\int}_{E^\text{th}_\text{IBD}}^{\infty}\,{dE_\nu}\,\sigma_\text{IBD}(E_\nu)\frac{dL_\nu}{dE_\nu}\left(t_R\right)\Theta(t_R)}{\dfrac{N_p}{4\pi D^2}\int_0^{\infty} dt\int_{E^\text{th}_\text{IBD}}^{\infty}\,{dE_\nu}\,\sigma_\text{IBD}(E_\nu)\frac{dL_\nu}{dE_\nu}\left(t\right)},
\end{align}
\end{widetext}
Note that the luminosity along LOS (i.e. $\theta=0$) is independent of $\phi$ (see Eq.~\eqref{eq:lum}).
It can be shown that since both signals originate through IBD interactions, integrating over sufficiently long observation times causes the integrals in the numerator and denominator to depend on the same combination of the neutrino spectrum. This important simplification allows us to define a clean theoretical expectation for this ratio in the case of isotropic emission,
\begin{align} \label{eq:R-iso}
\text{R}_{\textbf{iso}}^\text{theory}&
=\frac{\eta_\gamma\,A_{\text{eff}}^{511}}{2 \, N_p}\frac{Y_p}{Y_e\,\sigma_c}.
\end{align}
This result implies that the Hyper-K IBD event rate can be used to define a theoretical prediction for the total 511~keV photon count expected at the gamma-ray telescope,
\begin{equation} \label{eq:N-511}
    {N}_{\gamma}^\text{expected}
=\left(\frac{\eta_\gamma\,A_{\text{eff}}}{2 \, N_p}\frac{Y_p}{Y_e\,\sigma_c}\right) \int_0^{\infty} dt\,\mathcal{R}_\text{HK}(t).
\end{equation}
To aid an intuitive understanding of the effect of the anisotropy parameters on both the GRE signal and the Hyper-K event rate, we calculate the change in the ratio $\text{R}_{\textbf{iso}}^\text{theory}$ valid to first order in either $\epsilon_D$ or $\epsilon_Q$. 
\begin{align}
\frac{\text{R}_{\textbf{dipole}}^\text{theory}}{\text{R}_{\textbf{iso}}^\text{theory}}&
= \frac{1+{\epsilon_D\cos{\theta_0}/2}}{1+{\epsilon_D\cos{\theta_0}}} \simeq 1-\frac{\epsilon_D\cos{\theta_0}}{2},
\end{align}
where the numerator reflecting the total GRE rate in the dipole case follows directly from Eq.~\eqref{eq:NestimateDipole} and the denominator reflecting the total HK rate is obtained by simply setting $\theta=0$ in Eq.~\eqref{eq:lum} and combining it with Eq.~\eqref{eq:LnuExp1}. For the quadrupole case, the independence of the numerator from $\epsilon_Q$ follows directly from Eq.~\eqref{eq:NestimateQuadrupole}, and the denominator is obtained by combining Eq.~\eqref{eq:lum} with Eq.~\eqref{eq:Lnu2} at $\theta=0$.
\begin{align}
\frac{\text{R}_{\textbf{quadrupole}}^\text{theory}}{\text{R}_{\textbf{iso}}^\text{theory}}
&=\frac{1}{1-\frac{\,\epsilon_Q}{2}\left(1-3\cos^2{\theta_0}\right)} \nonumber \\ & \simeq 1+\frac{\,\epsilon_Q}{2}\left(1-3\cos^2{\theta_0}\right).
\end{align}
In the dipole case, the ratio acquires a correction linear in $\epsilon_D$, modulated by $\cos\theta_0$, reflecting the fact that the sensitivity to the dipole anisotropy depends directly on the orientation of the dipole axis relative to the line of sight to Earth. In the quadrupole case, the correction depends on $\epsilon_Q(1 - 3\cos^2\theta_0)$, which vanishes at the magic angle $\theta_0 = \arccos(1/\sqrt{3}) \approx 54.7^\circ$, implying that an observer located at this particular viewing angle would be completely insensitive to a quadrupolar emission asymmetry. 

By exploiting two key properties of the GRE flux signal evident in Fig.~\ref{fig:GREexp} --- namely, the rapid rise time and the subsequent plateau out to the characteristic timescale set by the stellar radius $R_\text{SN}/c$ --- we can improve the sensitivity of our method beyond a simple integrated event count by leveraging the time structure of the GRE signal itself in the case of anisotropic emission. To this end, we bin the observed GRE data in time intervals comparable to the characteristic decay timescale of the neutrino flux, which can be determined directly from the Hyper-K neutrino flux measurements. The expected GRE event count per time bin for dipole and quadrupole case for Rigel is shown in Fig.~\ref{fig:EventsRigel}. For the isotropic emission in both cases  ($\epsilon_{D/Q} = 0$), the expected event count is uniform across time bins, as shown by the blue curve, consistent with the flat plateau of the GRE signal observed after its rapid rise in Fig.~\ref{fig:GREexp_c}.

We also display the expected binned photon counts for non-zero anisotropy parameters $\epsilon_D$ (left panel) and $\epsilon_Q$ (right panel) in Fig.~\ref{fig:EventsRigel}. As discussed in Sec.~\ref{sec:GRE} and illustrated in Fig.~\ref{fig:GREAnisoRigel}, the GRE flux rises rapidly at early times for positive $\epsilon_D$ before asymptoting down to the isotropic plateau, while for negative $\epsilon_D$ the flux rises more gradually and only reaches the isotropic plateau at the characteristic timescale $R_\text{SN}/c$. This behavior is directly reflected in the binned photon count rates: the red curve ($\epsilon_D = +0.15$) shows an elevated event count at early times that decreases toward the isotropic average, while the green curve ($\epsilon_D = -0.15$) exhibits the opposite trend, with a suppressed early-time count that rises to meet the isotropic average at later times. Similarly for the expected binned photon counts for non-zero $\epsilon_Q$ (right panel) in Fig.\ref{fig:EventsRigel}, the behavior follows directly from the properties of its GRE flux as discussed in Sec.~\ref{sec:GRE} and illustrated in right panel of Fig.~\ref{fig:GREAnisoRigel}. 

An important subtlety concerns the theoretical expectation for the binned GRE event rate as inferred by a terrestrial observer from the Hyper-K measurements. Under the null hypothesis of isotropic emission, the Hyper-K-based prediction yields a flat binned GRE event rate. However, if the true SN neutrino emission is anisotropic, the Hyper-K event rate is itself modulated by the anisotropy parameter $\epsilon_{D/Q}$, which biases the observer's prediction for the GRE photon count. E.g. for positive $\epsilon_D$, the Hyper-K event rate is enhanced relative to the isotropic case, leading to an upward-biased prediction for the GRE photon count. However, since anisotropic emission with positive $\epsilon_D$ suppresses the neutrino flux at the larger polar angles that dominate the GRE signal at late times, the actual observed GRE event count will be statistically lower than this prediction. The opposite behavior holds for negative $\epsilon_D$. Since the sensitivity to the anisotropy parameter improves with the signal event count, the case of negative $\epsilon_D$ --- for which the observed GRE count exceeds the Hyper-K-based prediction --- requires a lower effective detector area to achieve the same level of detection sensitivity as the corresponding positive $\epsilon_D$ case of equal magnitude.

To assess the sensitivity of this method to the anisotropy parameter, we adopt a binned Poisson likelihood ratio test statistic~\cite{JUNO:2015zny}.  
\begin{widetext}
    \begin{equation}
    \Delta\chi^2= \min_{{\xi_\text{norm}}}\left[2\, \sum_{i} \left((1+\xi_\text{norm})\,\text{n}_{i}^\text{expected} -\text{n}_{i}^{\epsilon} +\text{n}_{i}^{\epsilon} \ln\left(\frac{\text{n}_{i}^{\epsilon} }{(1+\xi_\text{norm})\,\text{n}_{i}^\text{expected}}\right)\right)+\left(\frac{\xi_\text{norm}}{\sigma_\text{norm}}\right)^2\right],
\end{equation}
\end{widetext}
where $\text{n}_{i}^{\epsilon}$  denotes the event count in the $i$-th time bin for the assumed true anisotropic model and  $\text{n}_{i}^\text{expected}$ is the corresponding expected event count prediction from the measured neutrino flux at Hyper-K. Note that $\epsilon$ here refers to either $\epsilon_D$ or $\epsilon_Q$.  The minimization is performed over the normalization uncertainty $\xi_\text{norm}$ with prior width $\sigma_\text{norm}$, which encapsulates the dominant stellar systematic uncertainty.

The primary error in determining the normalization arises from the ratio $(Y_p/Y_e)$ of the stellar envelope composition. This ratio can directly be written in terms of the helium fraction at the stellar surface 
\begin{equation}
    \frac{Y_p}{Y_e}= \frac{n_H}{n_H+2\,n_\text{He}}
\end{equation}
where $n_H$ and $n_\text{He}$ are the number densities of hydrogen and helium respectively. For Rigel, Ref.~\cite{Przybilla:2005na} reported the photospheric helium abundance in terms of the number fraction $y=n_\text{He}/(n_H+n_\text{He})$. The authors infer $y$ from high-resolution optical spectroscopy by comparing the observed hydrogen and helium absorption-line profiles with spectra calculated using a model of Rigel's atmosphere. Helium abundance is constrained primarily by the weak He~I lines that are most sensitive to changes in $y$. This analysis gives $y=0.135\pm0.020$, resulting in a relative uncertainty of approximately $4\%$ for $({Y_p}/{Y_e})$. Rigel was subsequently reanalyzed in Ref.~\cite{Firnstein:2012aj} as part of a larger and more homogeneous sample of Galactic BA-type supergiants leading to a refined uncertainty of approximately $2.2\%$. However, the central values inferred from the two analyses differ by approximately $6\%$. We therefore adopt a nominal $5\%$ uncertainty on $(Y_p/Y_e)$ for Rigel.

Such precision is not expected to be generic for all nearby progenitor candidates. In particular, the much cooler Betelgeuse does not provide the same directly observable helium fraction. However in the event of a Betelgeuse supernova, observations of the shock-breakout and subsequent shock-cooling light curve, combined with immediate UV/optical spectroscopy and stellar-envelope modeling, could provide additional constraints on the He composition.

Given the more directly constrained composition and the comparatively small 511~keV background toward Rigel, we focus our primary sensitivity study on Rigel. We present our results in Fig.~\ref{fig:ChiSquaredRigel}, showing the $\Delta\chi^2$ as a function of the effective area $A_\text{eff}$ of the 511~keV gamma-ray detector for two values of the given anisotropy parameter $\epsilon_D$ (left panel) or $\epsilon_Q$ (right panel), assuming Rigel as the SN progenitor and for two different values of $\sigma_\text{norm}=\{5\%,\,10\%\}$ uncertainty on the flux normalization arising from the modeling of the stellar quantities $Y_p$ and $Y_e$. A first notable feature is that the required $A_\text{eff}$ for the green curve ($\epsilon_D = -0.15$) for given $\Delta\chi^2$ is smaller than that for the red curve ($\epsilon_D = +0.15$), consistent with the discussion above. Detecting a $15\%$ dipole anisotropy at $68\%$~C.L. requires $A_\text{eff} \sim 10^4~\text{cm}^2$ in both cases. At $90\%$~C.L., however, the required effective area for $\epsilon_D = -0.15$ increases to $\sim 3 \times 10^4~\text{cm}^2$, while for $\epsilon = +0.15$ it exceeds $4 \times 10^4~\text{cm}^2$. For quadrupole emission, anisotropy detection for $|\epsilon_Q|=0.2$ at $68\%$~C.L. requires $A_\text{eff} \sim 5 \times 10^3~\text{cm}^2$ in both cases. The required effective area increases to $\sim (2.0-2.5) \times 10^4~\text{cm}^2$ for detection at $90\%$~C.L. This demonstrates that for a given $A_\text{eff}$, our setup is significantly more sensitive to quadrupole anisotropy than to dipole anisotropy. The physical origin of this behavior is that the integrated GRE flux is insensitive to the quadrupole anisotropy parameter. More generally, we expect a similar distinction between odd and even multipoles for emission patterns described by $P_n(\cos\theta)$.

\begin{figure*}[htbp!]
    \centering \includegraphics[width=0.49\linewidth]{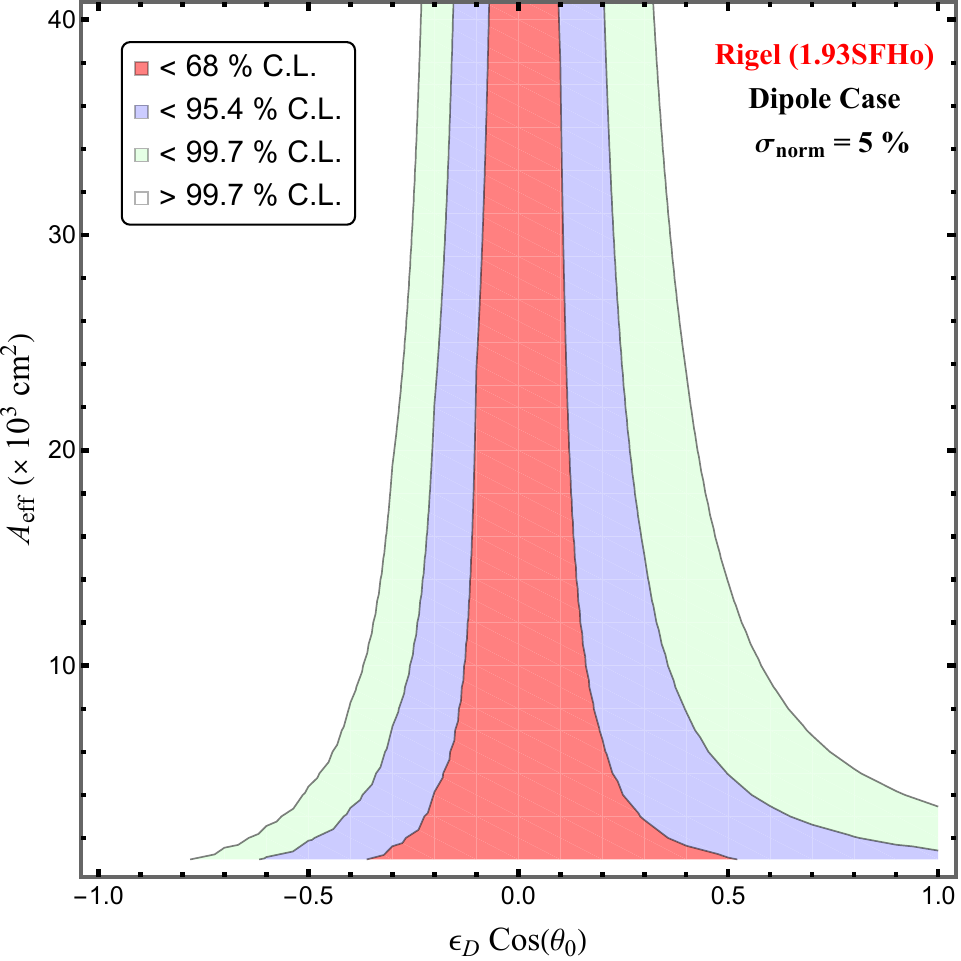}
    \includegraphics[width=0.49\linewidth]{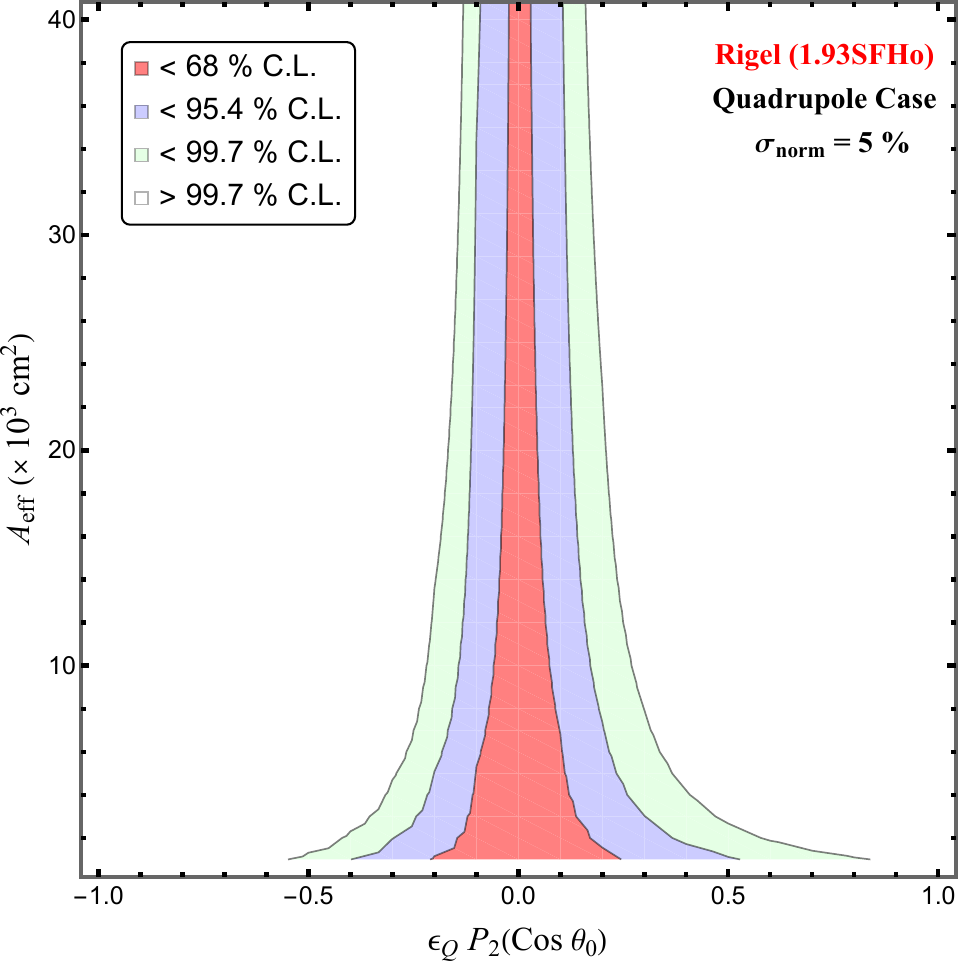}
    \caption{Contour plot of constant $\Delta\chi^2$ as a function of the gamma-ray detector effective area $A_\text{eff}$ and the effective anisotropy parameter, $\epsilon_D\,\cos\theta_0$ (left) and $\epsilon_Q\,P_2(\cos\theta_0)$ (right) for normalization uncertainty of $\sigma_\text{norm}=5\%$, assuming Rigel as the SN progenitor. The three contours correspond to 1, 2 and 3 $\sigma$ confidence levels.}
    \label{fig:FullChiSquaredScan}
\end{figure*}
For a rapidly rotating progenitor, the quadrupolar emission axis is set by the rotation axis itself: the persistent polar accretion downflows and the centrifugally induced oblate deformation of the proto-neutron star are both axisymmetric about the spin axis, and drive a dominant quadrupolar mode in the $\bar{\nu}_e$ luminosity~\cite{Walk:2019ier}. The inclination of this axis relative to the line of sight can then be constrained independently of the neutrino data, from the viewing-angle dependence of the gravitational-wave signal of a rapidly rotating progenitor~\cite{Takiwaki:2017tpe}, or from early-time spectropolarimetry of the explosion, which reveals the symmetry axis of the ejecta. Such an external determination of $\theta_0$ breaks the $\epsilon_Q$--$\theta_0$ degeneracy and allows the quadrupole amplitude to be extracted on its own. For the dipolar case, no comparable handle is available: LESA develops stochastically from convective fluctuations within the proto-neutron star, and its dipole direction migrates over the emission surface rather than locking to the rotation
axis~\cite{Tamborra:2014aua,Walk:2019miz}. 

We therefore quote dipole and quadrupole sensitivities in terms of an effective combination of the anisotropy parameter and the appropriate Legendre polynomial evaluated at $\cos\theta_0$, with the understanding that the measurement constrains the combination rather than $\epsilon_{D}$ or $\epsilon_{Q}$ alone. Our results are shown in Fig.~\ref{fig:FullChiSquaredScan} as a contour plot of the detector effective area $A_\text{eff}$ as a function of the above defined effective anisotropy parameter for both dipole (left panel) and quadrupole (right panel). We assume the normalization uncertainty from the stellar systematic is $\sigma_\text{norm}=5\%$. The three contours of the red, blue and green regions correspond to 1, 2 and 3 $\sigma$ confidence levels, respectively. For the dipole case, we can see that the sensitivity becomes systematics limited beyond $A_\text{eff}\sim 30\times10^3\text{ cm}^2$ for $-0.25\lesssim\epsilon_D\,\cos\theta_0\lesssim 0.36$. While for the quadrupole case, a large portion of the $\epsilon_Q\,P_2(\cos\theta_0)$ parameter space can already be probed with $A_\text{eff}\lesssim 20\times10^3\text{ cm}^2$. The sensitivity becomes systematics limited beyond this region for $-0.17\lesssim\epsilon_Q\,P_2(\cos\theta_0)\lesssim 0.21$. Notice that the sensitivities in both contour plots are not symmetric about zero, the reason for which was discussed in relation to Fig.~\ref{fig:ChiSquaredRigel}.

It is worth noting that for non-standard propagation scenarios, such as neutrino decay or flux attenuation, an analogous ratio can be defined with additional simplifications arising from the assumption of isotropic SN neutrino emission and time integration. In such cases, $\epsilon_\text{BSM}$ (defined in Eq.~\eqref{eq:LnuBSM}) is expected to be positive, since non-standard attenuation or decay reduces the $\bar{\nu}_e$ flux arriving at Earth relative to that inferred from the GRE signal, resulting in an observed GRE event rate that exceeds the prediction based on the Hyper-K $\bar{\nu}_e$ flux measurement. A negative value of $\epsilon_\text{BSM}$, while less natural, could arise for instance from the decay of heavy sterile neutrinos outside the SN envelope, producing active neutrino secondaries at higher energies; these would enhance the IBD event rate at Hyper-K without any corresponding enhancement of the GRE signal, thereby driving $\epsilon_\text{BSM} < 0$.

\section{BSM case study: invisible neutrino decay}
\label{sec:BSM}
The comparison developed in Sec.~\ref{sec:IBDratio} is not restricted to anisotropic neutrino emission. It can also probe new physics that modifies the neutrino flux during propagation from the stellar surface to Earth. In this case, the gamma-ray echo acts as a near-source measurement of the neutrino fluence, whereas Hyper-K measures the same burst after propagation over the full SN--Earth baseline. A discrepancy between these two measurements would therefore provide evidence for propagation-induced attenuation without requiring a theoretical prediction for the absolute supernova neutrino luminosity.

As a concrete example, we consider invisible neutrino decay. The purpose of this section is not to perform a detailed spectral fit, but to illustrate how the GRE--Hyper-K comparison can be translated into a sensitivity to the neutrino lifetime-to-mass ratio. We assume isotropic emission throughout this section. For this proof-of-principle study, we also assume that all three mass eigenstates have the same lifetime-to-mass ratio denoted by ${\tau}/{m}$
and decay exclusively into invisible final states. Therefore, the common survival probability is given by
\begin{equation}
S(E_\nu)=
\exp\left[
-\frac{D_{\mathrm{SN}}}
{c\,E_\nu\,(\tau/m)}
\right]\equiv (1 - \epsilon_{\text{dec}}).
\label{eq:decay_survival}
\end{equation}
where $\epsilon_{\text{dec}}$ is defined based on the notation of Eq.~\eqref{eq:LnuBSM}.
\begin{figure}[t]
    \centering
    \includegraphics[width=1.0\linewidth]{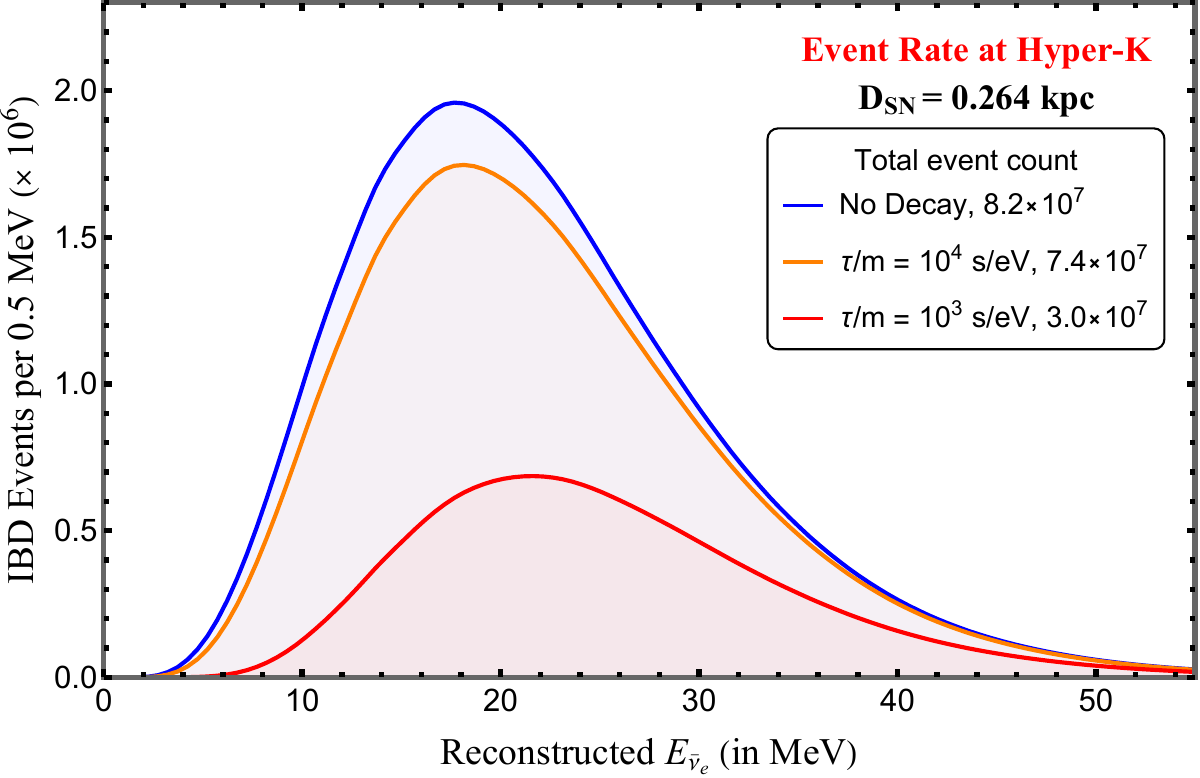}
    \caption{Reconstructed Hyper-K IBD spectrum for the simulated neutrino flux from $20\,M_\odot$ SN progenitor at a distance $D_{\mathrm{SN}}=264$ pc in the standard scenario and for two different values of invisible-decay lifetime-to-mass ratios.}
    \label{fig:hk_decay_spectrum}
\end{figure}
The invisible neutrino decay suppresses the Hyper-K event count while leaving the GRE flux unchanged, since the latter is produced in the stellar envelope before the neutrinos can undergo any substantial decay. We show the reconstructed energy spectrum for IBD events at Hyper-K for Rigel as the SN progenitor for three different cases (see fig.~\ref{fig:hk_decay_spectrum}), standard scenario with no decay (blue), for $\tau/m=10^4~\mathrm{s/eV}$ (orange) and $\tau/m=10^3~\mathrm{s/eV}$ (red). The figure clearly illustrates the energy-dependent suppression caused by the invisible decay.

Let $N_{\mathrm{HK}}=\int dt\,\mathcal{R}_{\mathrm{HK}}(t)$ denote the total Hyper-K count and the total GRE photon count given by $N_{\gamma}$. For the integrated-count comparison, we evaluate the survival probability at a representative burst energy $E_{\nu\star}$ and define $S_\star\equiv S(E_{\nu\star})$. The decay hypothesis then gives $N_{\mathrm{HK}}^{\mathrm{dec}}=S_\star N_{\mathrm{HK}}^{\mathrm{0}}$, while $N_{\gamma}^{\mathrm{dec}}=N_{\gamma}^{\mathrm{0}}$. Using these give
\begin{equation}
\frac{
N_{\gamma}^{\mathrm{dec}}/N_{\mathrm{HK}}^{\mathrm{dec}}
}{
N_{\gamma}^{\mathrm{0}}/N_{\mathrm{HK}}^{\mathrm{0}}
}
=
\frac{1}{S_\star}.
\label{eq:decay_count_ratio}
\end{equation}
The absolute supernova luminosity cancels from this comparison. This is a key distinction from neutrino-only analyses of SN1987A~\cite{Ivanez-Ballesteros:2023lqa}, in which the unattenuated fluence must be reconstructed using an assumed flux model for the supernova neutrino emission. The GRE instead provides an independent empirical normalization of the same IBD-weighted fluence before propagation losses occur. Invisible decay can therefore be tested through a mismatch between two observables, rather than through a deficit relative to an assumed source luminosity.

Since the Hyper-K event sample is much larger than the expected GRE photon sample, we neglect its statistical uncertainty. We denote by $\sigma_{\mathrm{GRE}}^{\mathrm{sys}}$ the fractional systematic uncertainty in predicting the GRE photon count from the Hyper-K measurement through Eq.~\eqref{eq:N-511}. This uncertainty is dominated by the stellar-envelope factors entering the Hyper-K-to-GRE conversion, in particular $Y_p/Y_e$. Combining this uncertainty in quadrature with the GRE photon-counting uncertainty, the total fractional uncertainty of the GRE-to-Hyper-K count ratio is
\begin{equation}
\begin{split}
\sigma_{\mathrm{rel}}
&\equiv
\frac{
\sigma\!\left(N_{\gamma}/N_{\mathrm{HK}}\right)
}{
N_{\gamma}/N_{\mathrm{HK}}
}
\\
&\simeq
\sqrt{
\frac{1}{N_{\gamma}}
+
\left(\sigma_{\mathrm{GRE}}^{\mathrm{sys}}\right)^2
}.
\end{split}
\label{eq:decay_fractional_uncertainty}
\end{equation}
The decay-induced enhancement is distinguishable from the standard expectation at the $n\sigma$ level when
\begin{equation}
\frac{1}{S_\star}-1
\simeq
n\,\sigma_{\mathrm{rel}}.
\label{eq:decay_sensitivity}
\end{equation}
We adopt $n=1.65$, corresponding to the conventional $90\%$ confidence criterion for a single parameter. For the Rigel benchmark, assuming $A_{\mathrm{eff}}=10^4~\mathrm{cm^2}$, we obtain $N_{\gamma}^0\simeq158$. Taking the systematic uncertainty on the GRE signal to be $\sigma_{\mathrm{GRE}}^{\mathrm{sys}}=5\%$ yields $\sigma_{\mathrm{rel}}\simeq0.094$, and consequently $S_\star\simeq0.866$. The comparison is therefore sensitive to an approximately $13\%$ suppression of the terrestrial IBD signal.

Solving Eq.~\eqref{eq:decay_sensitivity} for $S_\star$ and substituting the result into Eq.~\eqref{eq:decay_survival} gives
\begin{equation}
\frac{\tau}{m}
\gtrsim
\frac{D_{\mathrm{SN}}}
{c\,E_{\nu\star}\ln\left(1+n\,\sigma_{\mathrm{rel}}\right)}.
\label{eq:decay_lifetime_reach}
\end{equation}
For $D_{\mathrm{SN}}=0.264$ kpc, $E_{\nu\star}=15$ MeV and the Rigel uncertainty quoted above, this corresponds to
\begin{equation}
\frac{\tau}{m}
\gtrsim
1.3\times10^4~\mathrm{s/eV}.
\end{equation}
Increasing the effective area reduces the photon-counting contribution to $\sigma_{\mathrm{rel}}$, but it cannot reduce the systematic uncertainty in the Hyper-K-to-GRE conversion. In the limit $N_{\gamma}\rightarrow\infty$, one has $\sigma_{\mathrm{rel}}\rightarrow\sigma_{\mathrm{GRE}}^{\mathrm{sys}}$, and the lifetime reach at fixed source distance approaches
\begin{equation}
\left(\frac{\tau}{m}\right)_{\mathrm{max}}
\simeq
\frac{D_{\mathrm{SN}}}
{c\,E_{\nu\star}
\ln\left(1+n\,\sigma_{\mathrm{GRE}}^{\mathrm{sys}}\right)}.
\label{eq:decay_systematic_limit}
\end{equation}
For Rigel, $\sigma_{\mathrm{GRE}}^{\mathrm{sys}}=5\%$ and $n=1.65$, this gives
\begin{equation}
\left(\frac{\tau}{m}\right)_{\mathrm{max}}
\simeq
2.3\times10^4~\mathrm{s/eV}.
\end{equation}
Therefore, further improvement from a given progenitor requires reducing the stellar-envelope uncertainty in the Hyper-K-to-GRE conversion, rather than increasing the gamma-ray effective area alone.

For comparison, the recent SN1987A analysis of Ref.~\cite{Ivanez-Ballesteros:2023lqa} obtained $\tau/m\gtrsim1.2\times10^5~\mathrm{s/eV}$ at $90\%$ confidence for inverted mass ordering within a specific three-flavor decay scenario and an assumed family of supernova spectra. Although the benchmark reach obtained in our work is numerically weaker, it is based on a qualitatively different observable independent of the assumed SN neutrino spectra.

In a neutrino-only observation, propagation losses can be degenerate with a smaller intrinsic supernova neutrino luminosity. The GRE breaks this degeneracy by providing an empirical normalization of the neutrino fluence before propagation toward Earth. Invisible attenuation can then be identified through a mismatch between the GRE and Hyper-K event counts, rather than through comparison with an assumed absolute source luminosity. For a future Galactic supernova, this near--far consistency test would provide an independent constraint on invisible decay and a direct cross-check of the information extracted from the terrestrial neutrino spectrum.

\section{Conclusions}
\label{sec:conclusion}
We have proposed a novel method to experimentally probe neutrino emission anisotropy during a nearby galactic CCSN, by combining the gamma-ray echo signal with the $\bar{\nu}_e$ events detected at terrestrial neutrino observatories. Since both signals  
originate in $\bar \nu_e$ interactions via IBD, comparing them effectively provides a  
robust test of the angular distribution of the $\bar \nu_e$ emission. 
Indeed, while the GRE signal depends on the neutrino flux that illuminates half of the stellar surface (the one facing Earth, see Fig. \ref{fig:mainfigure}), the IBD event rate at Hyper-K measures the flux along the SN--Earth line of sight exclusively. By leveraging the distinct angular dependences of these two observables, we have demonstrated how both dipolar and quadrupolar emission anisotropies can be constrained, provided sufficiently large effective-area gamma-ray telescopes operating at 511~keV are available. Concretely, we find that, for a near-Earth supernova like Rigel, detecting a dipole anisotropy at the level of $\epsilon \sim 15\%$ at $68\%$~C.L. requires an effective detector area of $A_\text{eff} \sim 10^4~\text{cm}^2$. Therefore, our results contribute to motivating the development of next-generation MeV gamma-ray telescopes with effective areas on this scale. 

We have also discussed how the stellar envelope and a detector on Earth naturally act as a near--far detector setup over an astrophysical baseline, opening a complementary window on BSM physics such as invisible neutrino decay. In the latter case, the near--far normalization comparison exploits the large $D_\text{SN}/(c\,E_\nu)$ lever arm of Galactic baselines to place competitive constraints on the neutrino lifetime-to-mass ratio, even without significant spectral distortions. The full potential of the near--far detector concept, including a systematic exploration of a broader range of BSM scenarios, will be developed in detail in a companion follow-up work. 

In the event of a near-Earth supernova, the detection of neutrino emission anisotropy from the method proposed here would provide direct evidence of large-scale asymmetries in the CCSN engine, such as those predicted by LESA or SASI. The exact mechanism could be determined by examining complementary observables, like the frequency spectrum of the GW signal and the time evolution of the neutrino luminosity over the first $\sim 0.1$ s post-collapse, which could carry SASI signatures (see, e.g. \cite{Lin:2022jea}).  
A mismatch between near and far detector (in the sense outlined above) could also be attributable to propagation effects, and therefore could constitute a clear signature of new physics in the neutrino sector at astrophysical energy scales. The two interpretations, BSM physics or neutrino emission anisotropy, could be distinguished, at least in principle, from the shape of the luminosity curve of the GRE, which depends on the anisotropy at play (Fig. \ref{fig:GREAnisoRigel}) but is independent of BSM effects. 

Although we have outlined the potential and broad applicability of this method, its full realization requires the development of very large effective-area gamma-ray telescopes operating at 511~keV, which remains technologically challenging with current instrumentation. Nevertheless, dedicated 511~keV instruments achievable with present technology can reach effective areas of up to $\sim 10^4~\text{cm}^2$, which is precisely the benchmark we have adopted throughout this work, thus underscoring the importance of continued investment in next-generation MeV gamma-ray astronomy.

Several important extensions of the present work remain to be pursued. A particularly significant generalization will be the treatment of time-dependent anisotropy, which is a more realistic description of the dynamical CCSN environment. In addition, the case of a neutrino luminosity with both (simultaneous) dipole and quadrupole components would provide a more complete characterization of the emission geometry.
One could also include flavor oscillations inside the Earth, which have been neglected here. These affect neutrinos which traverse a substantial chord of the Earth before reaching the detector, and consist of additional oscillatory spectral modulations relative to the flux exiting the star. Since these effects depend on the detector's zenith angle to the source, a combined analysis of Hyper-K and IceCube (geographically well separated and hence exposed to different Earth-crossing baselines) offers a natural handle for disentangling them.

On the BSM side, the near--far framework outlined here can be applied to a variety of new physics scenarios beyond invisible decay, including neutrino--dark matter scattering. The case of pseudo-Dirac neutrinos will be explored in detail in a follow-up work. Finally, the role of IceCube as a high-statistics far detector has not been exploited in the present analysis and can be a useful future research direction. The IceCube effective volume for SN neutrino detection is directly proportional to the positron track length~\cite{IceCube:2023ogt} and hence to the positron energy, which renders the IceCube event rate sensitive to the third moment, $\langle E_\nu^3 \rangle$. This overall sensitivity to the neutrino energy spectrum may provide additional discriminating power between emission anisotropy and BSM propagation effects.

\textbf{Acknowledgements.}
The work of GC and CL is supported by the NSF Awards Number 2309973 and 2609687. GC and YP thank the Center for Theoretical Underground Physics and Related Areas
(CETUP* 2025 and CETUP* 2026) and the Institute for Underground Science at SURF for hospitality and for providing a stimulating environment, where a part of this work was done.

\bibliography{ref}

\end{document}